\documentstyle[epsfig]{aipproc}
\newcommand{\epsfigure}[2]{\epsfig{file=#1,width=#2}}

\begin{document}
\title{Storage Capacity of the Tilinglike Learning Algorithm}

\author{Arnaud Buhot$^*$ and Mirta B. Gordon$^{\dagger}$
\thanks{also at Centre National de la Recherche Scientifique}
}
\address{$^*$Theoretical Physics, University of Oxford, 
1 Keble Road, Oxford OX1 3NP, UK.\\
$^{\dagger}$SPSMS, DRFMC, CEA Grenoble, 17 av. des Martyrs,
38054 Grenoble Cedex 9, France.\\}

\maketitle

\begin{abstract}
The storage capacity of an incremental learning algorithm 
for the parity machine, the Tilinglike Learning Algorithm,
is analytically determined in the limit of a large number of 
hidden perceptrons. Different learning rules for the
simple perceptron are investigated. The usual Gardner-Derrida
one leads to a storage capacity close to the upper bound, 
which is independent of the learning algorithm considered.
\end{abstract}

\section{Introduction}

The storage capacity is one of the most 
important characteristics of a neural network. 
It is the maximal number of random 
input-output patterns per input entry that a network 
is able to correctly classify with probability one.
This quantity is independent of the algorithm used to 
learn the weights of the network; it only 
depends on its architecture. We will refer to it as the
{\it architecture storage capacity} $\alpha_c^{arch}$ 
in the following, to distinguish it from the 
algorithm-dependent one, hereafter 
called {\it algorithm storage capacity}, $\alpha_c^{alg}$.

The simplest neural network, the perceptron, has 
the inputs directly connected to the output. 
Geometrical arguments~\cite{Cover} and a 
statistical mechanics calculation~\cite{Gardner} 
determined that $\alpha_c^{arch} = 2$. 
Several perceptron learning algorithms, like the 
Adatron~\cite{AbKe,AnBi}, are 
known to achieve such storage capacity. 
  
The capacity can be increased using networks with  
more complicated architectures. 
The next one on increasing complexity is the 
extensively studied~\cite{WRB}
monolayer perceptron (MLP), which has 
$k$ ``hidden'' perceptrons connected 
to the output unit. As a MLP 
can store any function of its inputs, 
provided that the number of hidden units 
is adequate, it is not worth to consider 
more complex architectures for the storage problem.  

Given an input pattern, the hidden 
units' states define a $k$-dimensional 
vector, the pattern's {\it internal 
representation} (IR). The network's output is 
a function of the IR. In the following, we consider 
binary neurons, of states $\pm 1$, and 
focus on the parity machine, whose output is the 
product of the $k$ components of the IR. 

The main problem when training MLPs is that 
the IRs of the input patterns 
are unknown. It has been proposed~\cite{BiOp} 
to build the hidden layer using a 
constructive procedure, called Tilinglike 
Learning Algorithm (TLA), in which the hidden perceptrons 
are included one after the other and 
trained to correct the learning errors of the 
preceding unit. As each unit can at least correct 
one error, convergence is ensured~\cite{BiOp}. It is 
straightforward to show that the TLA generates 
a parity machine~\cite{BiOp}, but 
the number of included hidden units 
depends crucially on the performance of 
the perceptron learning algorithm used 
to train them.

Geometric arguments~\cite{MiDu}
and a statistical mechanics replica calculation~\cite{XKO}, 
showed that the architecture storage capacity of 
the parity machine in the limit of a large number 
of hidden units $k$ is $\alpha_c^{arch} (k) \simeq k 
\ln k/\ln 2 \simeq 1.44 \, k \ln k$. However, it was 
not clear whether this storage capacity could be 
actually achieved with a learning algorithm. 
In this paper, we show that the Tilinglike 
Learning Algorithm (TLA) can reach a storage 
capacity close to the architecture storage capacity, 
provided that the hidden perceptrons are trained with an 
appropriate learning algorithm. In section \ref{sec.TLA} we 
describe more precisely the setting and the TLA. The 
analytical expression of the algorithm storage 
capacity in the limit of large $k$ is determined 
in section \ref{sec.alpha}, where 
we show that the learning algorithm 
used to train the hidden units must satisfy 
stringent condition for the TLA to converge 
with a finite number of hidden units. The results 
presented in section \ref{sec.algo} show that 
these conditions rule out some perceptron 
learning algorithms, like the Adatron. 
The conclusions are presented in section \ref{sec.concl}. 

\section{The Tilinglike Learning Algorithm}
\label{sec.TLA}

Let us assume a training set ${\cal L}_{\alpha} = \left\{ 
{\bf x}_\mu, \tau_{\mu} \right\}_{\mu= 1, \cdots, P}$ of
$P = \alpha N$ input-output patterns. The inputs ${\bf x}_\mu$ 
are random gaussian $N$-dimensional vectors with zero mean and 
unit variance in each direction. The corresponding outputs 
$\tau_{\mu}=\pm1$ are the learning targets. Their values 
$\tau$ are randomly selected with probability:
\begin{equation}
\label{proba}
P(\tau;\varepsilon) = \varepsilon \, \delta (\tau + 1) 
+ (1 - \varepsilon) \, \delta (\tau - 1), 
\end{equation}
\noindent with $\varepsilon = 1/2$. The r\^ole of the 
bias $\varepsilon$ introduced in (\ref{proba}) will become 
clear in the following. The probability of the targets 
$\tau_{\mu}$ is unbiased.  
 
The TLA constructs the parity machine by including 
successive perceptrons in the hidden layer. Each unit 
is connected to the input $\bf x$ through 
weights $\{ {\bf J},\theta \}=(J_1, \dots, J_N,\theta)$ where 
$\theta$ is a threshold. The inputs are classified through 
$\sigma = {\rm sign} ( {\bf J} \cdot {\bf x} - \theta )$. 
Thus, a perceptron separates linearly the input space with 
a hyperplane orthogonal to ${\bf J}$ 
(we assume ${\bf J} \cdot {\bf J} = 1$) 
at a distance $\theta$ to the origin. The weights and 
the threshold are learned through the minimization of a 
cost function:
\begin{equation}
\label{cost}
E \left( \left\{ {\bf J},\theta \right\}; {\cal L}_{\alpha}
\right) = \sum_{\mu = 1}^{P} V \left(\tau_{\mu} ({\bf J} 
\cdot {\bf x}_\mu - \theta) \right).
\end{equation}
where the potential $V(\lambda)$ is the contribution 
of each pattern to the cost, and $|\lambda|$ is the 
distance of the pattern to the hyperplane.  

Within the TLA heuristics, the first perceptron $k=1$ 
is trained to learn targets $\tau_{\mu}^1=\tau_{\mu}$. 
After learning, its weights are $\{ {\bf J}^1, \theta^1 \}$; 
its training error is $\varepsilon_t^1 = (1/P) 
\sum_{\mu} \Theta (-\sigma^1_{\mu} \tau^1_{\mu})$ 
where $\Theta(x)$ is the Heaviside function 
and $\sigma^1_{\mu}$ the perceptron's output to 
pattern ${\mu}$. If $\varepsilon_t^1 = 0$, the 
training set is correctly classified; the TLA 
stops with only one simple perceptron. Otherwise, 
a new perceptron is introduced. The successive 
perceptrons $i$ are trained to learn training 
sets ${\cal L}_{\alpha}(i) = \{ {\bf x}_{\mu},
\tau^i_{\mu} \}$ with targets $\tau^i_{\mu} = 
\tau^{i-1}_{\mu} \sigma^{i-1}_{\mu}$, that is, 
$\tau^i_{\mu}= 1$ if the pattern $\mu$ is 
correctly classified by the previous perceptron 
and $\tau^i_{\mu} = - 1$ otherwise. 
If the perceptron learning algorithm is 
correctly chosen it can be shown that the 
successive training errors $\varepsilon_t^i$ are strictly 
decreasing~\cite{BiOp,Gordon}. Thus, the 
TLA procedure necessarily converges to a MLP 
with $k$ units, where the $k^{th}$ perceptron 
is the first one to meet the condition 
$\varepsilon_t^k=0$. Then, the product 
$\sigma_{\mu} = \sigma_{\mu}^{1} \cdots \sigma_{\mu}^k = \tau_{\mu}$
gives the correct output to the patterns of the training set 
${\cal L}_{\alpha}$. 

\section{Storage Capacity}
\label{sec.alpha}

The algorithm storage capacity of the TLA, 
$\alpha_c^{alg}(k)$, is simply the inverse 
function of $k(\alpha)$, the average number 
of perceptrons typically included by the 
TLA when the training set has a size $\alpha = P/N$.
In order to determine $k(\alpha)$, consider 
the $i^{th}$ hidden unit : The probability 
of its targets $\tau_\mu^i$ depends on the training error 
$\varepsilon_t^{i-1}$ of the previous perceptron.
Although there exist some correlations between the 
outputs $\tau_{\mu}^i$, due to the correlations in the 
weights of the successive perceptrons, in the 
limit of a large training sets 
($\alpha \rightarrow \infty$) they may be neglected~\cite{WeSa}.
Thus, we may assume that the targets $\tau_{\mu}^i$ are 
independently drawn with probability (\ref{proba}), with 
a bias $\varepsilon_t^{i-1}$. Then,
the successive training errors satisfy a simple 
recursive relation~\cite{WeSa,BuGo,Buhot}:
\begin{equation}
\varepsilon_t^{i} = {\cal E}_t \left(\alpha, 
\varepsilon_t^{i-1} \right)
\end{equation}
\noindent where ${\cal E}_t (\alpha, \varepsilon_t^{i-1})$ is the 
training error of a simple perceptron trained with a 
training set of size $\alpha$ and biased targets  
$\tau_{\mu}^{i}$ drawn with a probability $P(\tau_{\mu}^i;
\varepsilon_t^{i-1})$ given by (\ref{proba}).
The number $k$ of perceptrons necessary to correctly classify
the initial training set satisfies~\cite{BuGo,Buhot}:
\begin{equation}
\circ_{k} f_{\alpha} (1/2) = \underbrace{f_{\alpha}
 \circ \cdots \circ f_{\alpha}}_{k \
{\rm times}} \ (1/2) = 0
\end{equation}
\noindent where $f_{\alpha} (\varepsilon)$ stands for
${\cal E}_t (\alpha,\varepsilon)$ and the symbol $\circ$
for the composition of functions.
In the limit of a large $\alpha$, the training error
${\cal E}_t (\alpha,\varepsilon)$ is close to $\varepsilon$
and the number of simple perceptrons $k(\alpha)$ is large.
It is thus possible to use the continuum limit:
\begin{equation}
\varepsilon_t^{i+1} - \varepsilon_t^i \simeq
\frac{1}{k} \frac{d\varepsilon}{dx} = \varepsilon - {\cal E}_t 
(\alpha,\varepsilon)
\end{equation}
\noindent with $x = i/k$. After integration of this differential
equation we obtain~\cite{BuGo,Buhot} the typical number of hidden units introduced by the TLA:
\begin{equation}
k(\alpha) = \int_{0}^{1/2} \frac{d\varepsilon}{\varepsilon - 
{\cal E}_t (\alpha, \varepsilon)}.
\end{equation}
\noindent $k(\alpha)$ depends on the specific cost 
function (\ref{cost}) used to train the perceptrons 
through ${\cal E}_t (\alpha, \varepsilon)$, which is 
the training error of a simple perceptron learning 
a training set with biased targets.
\begin{figure}[t] 
\begin{center}
\epsfigure{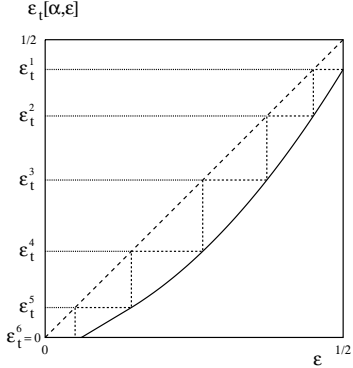}{5cm}
\end{center}
\vspace{10pt}
\caption{Successive training errors $\varepsilon_t^i$.
The full curve corresponds to ${\cal E}_t (\alpha, \varepsilon)$.
In this case six perceptrons are necessary for the convergence
of the Tilinglike Learning Algorithm.}
\end{figure}  

\section{Results for different learning potentials}
\label{sec.algo}

In this section we determine the perceptron's 
training error ${\cal E}_t (\alpha, \varepsilon)$ 
for biased training sets, in the 
thermodynamic limit ($N \rightarrow \infty$ with 
$\alpha = P/N$ fixed). For different 
possible choices of the potential $V$ in (\ref{cost}), 
we deduce the number of hidden units, $k(\alpha;V)$ 
and the algorithm storage capacity of the TLA.
Only the main results are presented here, the interested reader 
can find the details in~\cite{BuGo,Buhot}.

\subsubsection{Adatron cost function}

The Adatron potential is $V(\lambda) = 
(\kappa - \lambda)^2 \Theta
(\kappa - \lambda)$, where $\kappa$ is a positive 
parameter called stability. All the patterns with 
negative $\lambda$ and those with $\lambda>0$ but closer 
than $\kappa$ to the hyperplane (which are correctly 
classified) contribute to 
the cost. The training error 
${\cal E}_t (\alpha, \varepsilon)$ is obtained 
through a replica calculation assuming 
replica symmetry (RS), which can be shown 
to hold for all $\alpha$. It turns out that for
fixed $\kappa$, in the limit of large 
$\alpha$, ${\cal E}_t (\alpha, \varepsilon) > 
\varepsilon$. As a consequence, the constraint 
that the successive training errors are strictly 
decreasing, necessary for the 
convergence of the TLA, is not 
satisfied. This problem can be circumvented 
at the price of considering $\kappa$ as 
a free parameter, and minimizing the training 
error with respect to it. In that case, ${\cal E}_t 
(\alpha, \varepsilon) < \varepsilon$ and  
the algorithm storage capacity is 
$\alpha_c^{alg} (k, Adatron) \simeq 4.55 \, \ln k$ 
in the limit of large $\alpha$.
 
\subsubsection{Gardner-Derrida cost function}

The potential of the Gardner-Derrida (GD) cost 
function~\cite{GaDe} is $V(\lambda) = 
\Theta (\kappa- \lambda)$. The hypothesis of 
RS is incorrect for this potential, and the 
obtained value of ${\cal E}_t$ is a lower 
bound to the true training error. Consequently, 
the replica calculation allows only to determine 
an upper bound to $\alpha_c^{alg}(k)$. If $\kappa=0$, the 
cost is nothing else but the number of misclassified 
patterns. It gives the lowest bound to ${\cal E}_t$.  
In the limit of large training set size $\alpha$, 
we obtain:
\begin{equation}
{\cal E}_t (\alpha,\varepsilon) \simeq
\varepsilon - \frac{2}{a^2(\varepsilon)} \, \frac{\ln \alpha}{\alpha},
\end{equation}
\noindent where $a(\varepsilon)$ satisfies 
$\varepsilon = [e^{a} ( 1-a ) -
1]/[2 \, (\cosh a - a \sinh a - 1)]$.
This leads to $k(\alpha,GD) \simeq 0.475 \, \alpha/\ln 
\alpha$ and $\alpha_{c}^{alg} (k,GD) \simeq 2.11 \, k \ln k$,
larger than $\alpha_c^{arch}(k)$, probably 
due to the failure of the RS hypothesis.  

In order to obtain a lower bound to $\alpha_c^{arch}(k)$
we used the Kuhn-Tucker cavity method~\cite{GeKr,BuGo,Buhot}, 
which gives an upper bound to ${\cal E}_t$. As a result 
of both calculations, we can bound $\alpha_c^{alg}(k,GD)$:
\begin{equation}
0.924 \, k \leq \alpha_c^{alg} (k,GD) \leq 2.11 \, k \ln k.
\end{equation}
On view of this result, we expect that the algorithm storage 
capacity of the TLA behaves like $k (\ln k)^{\nu}$ with 
$0 \leq \nu \leq 1$. A calculation 
with one step of replica symmetry breaking  would give 
an estimate of the exponent $\nu$. Since the RS solution 
gives a better approximation of the training 
error than the Kuhn-Tucker cavity method, we expect 
the exponent $\nu$ to be close to $1$, leading to an 
algorithm storage capacity close
to the architecture's capacity. This result shows 
that the TLA  may build a nearly optimal network.  

In order to improve the robustness against noise in the data,
it is usually useful to impose some finite stability $\kappa$
to the patterns. The corresponding GD potential is 
$V(\lambda)=\Theta(\kappa-\lambda)$. As with $\kappa=0$, 
here also the RS solution is unstable. The bounds 
on $\alpha_c^{alg} (k, \kappa, GD)$ 
deduced from the results for ${\cal E}_t$ obtained 
with the RS hypothesis and the Kuhn-Tucker 
cavity method~\cite{BuGo,Buhot}, give: 
\begin{equation}
\frac{k}{\kappa \sqrt{2 \ln k}} \leq \alpha_c^{alg} (k, 
\kappa,GD) \leq \frac{k}{2 \kappa^2}.
\end{equation}
Strikingly, imposing a finite stability $\kappa$ has an 
important effect on the algorithm storage capacity,
which in this case behaves as $k (\ln k)^{\nu}$ with $-1/2 \leq \nu 
\leq 0$. The prefactors of the bounds of $\alpha_c^{alg} (k,
\kappa,GD)$ are $\kappa$-dependent and they both diverge
for $\kappa \rightarrow 0$. The exponent $\nu$ is independent 
of $\kappa$ for finite $\kappa$ but differs from the one 
corresponding to $\kappa = 0$.

\section{Conclusion}
\label{sec.concl}

We determined analytically the 
storage capacity of the Tilinglike Learning 
Algorithm for the parity machine, a constructive 
procedure generating a monolayer perceptron 
of binary hidden units. A training set of 
input-output examples is used to determine 
the number of hidden units, which are introduced 
one after the other. These are simple perceptrons that have to 
learn their weights using increasingly biased 
target distributions.

We have shown that the storage capacity of the 
TLA depends crucially on the learning errors 
of the successively introduced perceptrons. 
The properties of the algorithm used to train 
the latter have thus dramatic consequences on the 
size of the hidden layer generated by the TLA, 
and may even hinder the convergence to a finite 
size network. This arises, in particular, if the perceptrons 
are trained with the Adatron algorithm unless 
the stability is adapted to the successive targets' biases. 

The smallest network is obtained using the Gardner-Derrida 
cost function with vanishing stability, 
which corresponds to minimizing the training error. 
Based on the results obtained within the replica symmetry 
hypothesis, and those using the Kuhn-Tucker cavity method, 
we expect a supra-linear storage capacity
$\alpha_c^{alg} (k,GD) \simeq k (\ln k)^{\nu}$ 
with $\nu > 0$, very close to the theoretical capacity 
corresponding to the architecture considered.

\vskip 0.4cm
The work of AB is supported by the Marie Curie Fellowship
HPMF-CT-1999-00328.

\end{document}